\documentclass[reprint, aps, superscriptaddress,preprintnumbers,showpacs]{revtex4-1}
\usepackage{amsmath, amsfonts}

\usepackage{xspace}
\usepackage{caption}
\usepackage{float}
\usepackage{amssymb}
\usepackage{color}
\usepackage[usenames,dvipsnames]{xcolor}

 \usepackage{algpseudocode} 
\usepackage{hyperref}
\usepackage{amsthm}
\usepackage{graphicx}
\usepackage{tipa}
\usepackage{dcolumn}
\usepackage{bm}
\usepackage{subfigure}
\usepackage{amssymb}
\usepackage{verbatim}
\usepackage{amscd}
\usepackage{amssymb}
\usepackage{setspace}
\usepackage[]{graphicx}        
\usepackage{amsthm}
\usepackage{natbib}
\usepackage{enumerate}
\usepackage{afterpage}
\usepackage{placeins}
\newcommand{\quotes}[1]{``#1''}
\newcommand{\expt}[1]{\left \langle #1 \right \rangle}

\definecolor{dark-gray}{gray}{0.3}

\newcommand{\paren}[1]{\left ( #1 \right )}
\newcommand{\atfixed}[1]{\left. #1 \right |}

\newcommand{\abs}[1]{\left | #1 \right |}

\newcommand{\bracket}[1]{{\left[ #1 \right ]}}

\newcommand{\myequation}[1]{\begin{eqnarray} #1 \end{eqnarray}}
\newcommand{\myequationn}[1]{\begin{align*} #1 \end{align*}}

\newcommand{\myeqref}[1]{{Eq. \ref{#1}}}
\newcommand{\appendref}[1]{{Append. \ref{#1}}}
\newcommand{\figref}[1]{{Fig. \ref{#1}}}

\def \voltage{\ensuremath{V}}
\def \grav{\ensuremath{g} \xspace}
\def \resist{\ensuremath{\Omega} \xspace}
\def \conduct{\ensuremath{\kappa} \xspace}
\def \curr{\ensuremath{I} \xspace}
\def \dens{\ensuremath{\rho} \xspace}
\def \meandens{\ensuremath{\bar{\dens}} \xspace}

\def \width{\ensuremath{\mathbf{L}_{x}} \xspace}
\def \height{\ensuremath{\mathbf{L}_{z}} \xspace}

\def \removcoeff{\ensuremath{\alpha_{\text{R}}}\xspace}
\def \addcoeff{\ensuremath{\alpha_{\text{A}}}\xspace}
\def \removfunct{\ensuremath{{r}}\xspace}
\def \addfunct{\ensuremath{{a}}\xspace}

\def \posit{\ensuremath{{\mathbf{r}}} \xspace}
\def \wavevec{\ensuremath{\mathbf{k}}\xspace}

\newcommand{\horizontality}[1]{ \frac{#1_{x}^{2}}{\abs{#1}^{2}} }
\newcommand{\horizprime}[1]{ \frac{\paren{#1_{x}}^{2}}{\abs{#1}^{2}} }
\newcommand{\diagonality}[1]{ \frac{#1_{x}#1_{z}}{\abs{#1}^{2}} }

\def \diffus{\ensuremath{\sigma} \xspace}
\def \noise{\ensuremath{D} \xspace}

\def \meanremov{\ensuremath{\mathcal{R}} \xspace}

\def \meanadd{\ensuremath{\mathcal{A}} \xspace}
\def \meanflow{\ensuremath{\textbf{J}} \xspace}
\def \chempot{\ensuremath{\mathcal{N}} \xspace}
\def \timederiv{\ensuremath{\mathcal{T}} \xspace}

\def \syssizesymbol{\ensuremath{\mathbf{L}} \xspace}
\def \picsyssize{24 \xspace}
\def \simsyssize{40}

\begin{document}

\title{
Feedback induced phase transitions in active porous media 
}
\author{Samuel A. Ocko}
\affiliation{Department of  Physics, Massachusetts Institute of Technology, Cambridge, Massachusetts 02139, USA}

\author{L. Mahadevan}
\affiliation{School of Engineering and Applied Sciences, Department of Physics, Harvard University, Cambridge, Massachusetts 02138, USA}
\date{\today} 

\begin{abstract}

Flow through passive porous media is typically described in terms of a linear theory relating current fluxes and driving forces, in the presence of a prescribed heterogeneous permeability. However, many porous systems such as glacial drainage networks, erosional river bed networks, vascular networks, social insect swarms and animal architectures such as termite mounds are continuously remodeled by the flow and thence modify the flow, i.e. they are active. Here we consider a minimal model for an active porous medium where flow and resistance are coupled to each other.  Using numerical simulations, we show that this results in both channelization and wall-building transitions depending on the form of the feedback. A continuum model allows us to understand the qualitative features of the resulting phase diagram, and suggests ways to realize complex architectures using simple rules in engineered systems.
\end{abstract}

\maketitle
\emph{Introduction:} 
%
Transport through porous media is important in many problems in physics, biology, geology, and engineering. While most study is limited to transport through a \emph{static} medium, transport can often feedback to modify the medium itself. These \emph{active} porous media co-evolve with the transport through them.  

Examples of active porous media abound. River networks are formed through the interplay of erosion, transport, and deposition\cite{schorghofer2004spontaneous, mahadevan2012flow}, lightning results from the interplay of conduction and dielectric breakdown, and electrical fuses are engineered to break down above a critical current\cite{Duxbury:1987dr, Zapperi:1997er}. All life exists in a world of gradients and physical flows, and biological systems often arrange matter through feedback mechanisms to control transport at the cellular\cite{Tero:2010bx, Alim13082013, Nakagaki:2007wg, HeatonGrowth,Heaton201212}, organismal\cite{Pries2009, Masumura:2009to, Song:2011wq, leNoble:2005fu, leNoble:2005fu, Szczerba:2005hz, kamiya1984adaptive}, and societal\cite{Turner2000uy, Starks:1999es, Turner:2010hn, Heinrich:1981ws, jost2007interplay, howse1966air} level. Specific examples {of active porous media in biology} include network formation of slime molds, formation and remodeling of vascular networks, and the mound and wax architectures of social insects.

Elements universal across these active porous media are conservation of flow, feedback from transport and stochasticity.  In drainage networks, erosion increases with current while deposition decreases, while fuses are more likely to break in high current, and dielectric breakdown is enhanced by large currents and fields. In biology, ants have been observed to remove corpses from high-wind areas and place them in low-wind areas \cite{jost2007interplay}, while termites are known to respond to mound damage by building in response to {air flows, humidity, and olfactory cues} \cite{Turner:2010hn, howse1966air}. 
A minimal distillation of the common elements in these different active porous media corresponds to a stochastically evolving network driven uniformly by fluxes and forces at the boundary due to pressure, voltage, or concentration gradients. {Since problems involving steady state diffusion of heat, concentration gradients, or flow through a porous material are mathematically  analogous to current flow through electrical circuits, we will use the language of circuit theory from now on.}

\emph{Transport Laws:}  We focus on a translationally symmetric case with periodic boundary conditions and a uniform driving voltage in the vertical $\hat{z}$ direction. The vertices are arranged in a square network, with the current between neighboring vertices given by
\myequation{\curr_{ij} = \frac{1}{\resist_{ij}} \paren{\bracket{\voltage_ {i} - \voltage_ {j}} + \grav \hat{z} \cdot \hat{\mathbf{r}}_{ij}}, \qquad \sum_{j} \curr_{ij} = 0.
 \label{eq:discrete_circuit}}
Each vertex $i$ either contains a particle $(\dens_{i} = 1)$ or is empty$(\dens_{i} = 0)$, with the resistance between full vertices higher than the resistance between empty vertices; $\resist_{ij} = 1 + \Delta \resist \paren{\dens_{i} + \dens_{j}}/2$. 

\emph{Activity Rules:}
Particles are removed from their vertices at a \emph{current-dependent} rate proportional to $\removfunct(v_{i})$, where $v_i = \sqrt{1/2 \cdot \sum_{j} \curr_{ij}^{2}}$ is the current through the cell \footnote{This particular expression for $v$  chosen for rotational invariance}. They are then are added to an empty vertex  with probability proportional to $\addfunct(v_{j})$, leading to to a simple algorithm for evolution of the medium\cite{Gillespie:2007bx}:
%
\begin{enumerate}\itemsep0pt \parskip-1pt \parsep-4pt \label{markov_dynamics}
\item Remove a particle from filled vertex  $i$ randomly selected with probability proportional to $\removfunct(v_{i})$.
\item Solve for the new current through the network.
\item Add the particle to an empty vertex $j$ randomly selected with probability proportional to $\addfunct(v_{j})$.
\item Solve for the new current through the network.
\end{enumerate}
%
We emphasize that every step conserves particle number, but the movement of particles is nonlocal in that the distance between vertices $i,j$ may be arbitrarily large \footnote{A related model involving local movement gives very similar behavior (\appendref{app:LocalDynamics})}. Note that the system lacks detailed balance and thus we cannot write down a free energy functional associated with the dynamics (\figref{fig:NonEqAndFluctCartoon}). 

{Since the addition and removal rates can either increase or decrease with local current in the active systems described earlier, we explore this range of possibilities in terms of two parameters $\removcoeff$, $\addcoeff$. For the removal process, we choose $\removfunct(v_{i} )= {v_{i}}^{-\removcoeff}$; at positive \removcoeff, particles in high current will be less likely to be removed(current seeking), and vice versa. For the addition process, we choose $\addfunct(v_{j}) = {v_{j}}^{\addcoeff}$; at positive \addcoeff, empty vertices with high current will likely be filled(current seeking) and vice versa. 
\begin{figure*}[h!]
\vspace{-4 mm}

For simplicity, we have chosen our functional forms such that in any circuit, $\removfunct(v_{i}) \propto \grav^{-\removcoeff}$, $\addfunct(v_{j}) \propto \grav^{\addcoeff}$. Because only the \emph{ratios} of currents are important, we may set this system of equations to be dimensionless by making the substitutions:
$\resist_{0} \to 1$ , $\Delta \resist \to \Delta \resist/\resist_{0}$,  $\grav \to 1$. This gives  four dimensionless parameters: $\Delta \resist, \removcoeff, \addcoeff, \meandens$.  We want the difference between filled and unfilled vertices to be large; here we choose $\Delta \resist  = 19$ such that the resistance between filled vertices is 20 times higher than empty vertices. 

\vspace{4 mm}
   \includegraphics[width=.99\textwidth]{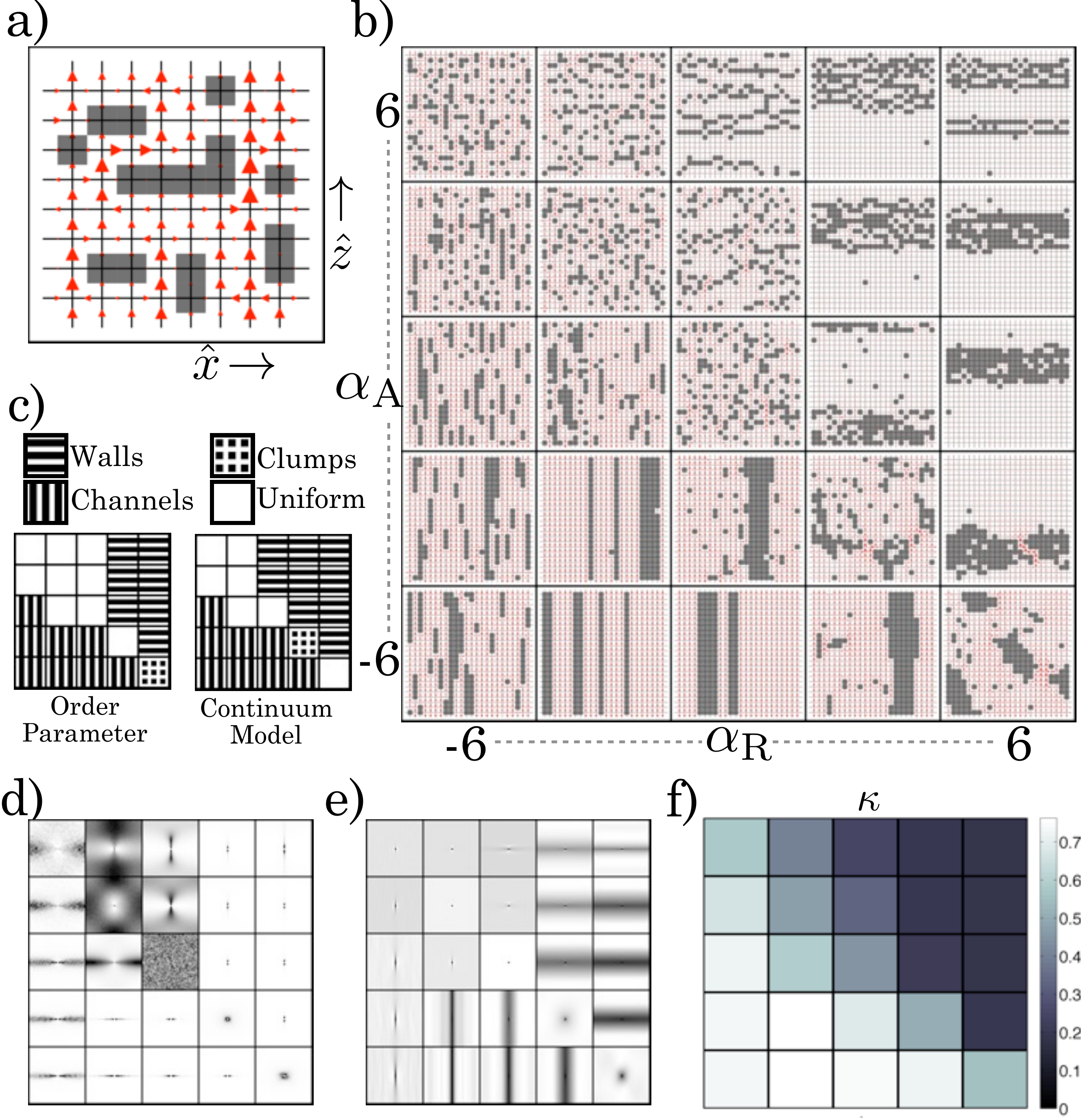} 
   \caption{a) Example system for $(\addcoeff, \removcoeff) = (0, 0)$. Filled vertices are covered by gray squares, unfilled vertices are not. Current is being driven in the upwards $\hat{z}$ direction, direction and magnitude of current between neighboring vertices is indicated by red arrows(color online).  
   b) Phase diagram for $\bar{\dens} = 1/4$. Each individual box represents a single system that has equilibrated for a particular $(\addcoeff, \removcoeff)$, where $\addcoeff, \removcoeff$ have values $(-6, 3, 0, 3, 6)$. At positive $\removcoeff$, the system has formed thick walls; negative $\removcoeff$ gives thin channels. At positive $\addcoeff$, the system has formed a series of thin walls; negative $\addcoeff$ gives thick channels. At positive $\removcoeff$, negative $\addcoeff$, a phase separation occurs at both orientations, as the system forms a set of {clumps}. 
   c) Observed phase transitions though observation of order parameters on ensemble(see text) vs. predictions of continuum model(see text).
 d)  Fourier transform $\expt{\tilde{\dens}(\wavevec)^{2}}$ (high amplitudes in dark, $\wavevec = 0$ at center). e) Two-point correlation $\expt{\paren{\dens(0) - \bar{\dens}} \paren{ \dens(\vec{\mathbf{r}})  -\bar{\dens} }}$  (positive correlations in dark, $\posit = 0$ at center).
    f)Contour plot of conductivity of the medium as a function of $\removcoeff, \addcoeff$.
Note that grids c)-f) use the same range of $\addcoeff, \removcoeff$ as grid b). Grids c)-f) use a system size of $\simsyssize \times \simsyssize$, while grid b) uses a smaller system size of $\picsyssize \times \picsyssize$ to aid in visualization. 
   }
   \label{fig:mega_picture}
\end{figure*}

\clearpage

\emph{Simulations:}
{For each $\removcoeff, \ \addcoeff, \ \meandens$, we start the system at a uniform density $\meandens$ and evolve it so that each particle or hole moves an average of one thousand times. This procedure is repeated twenty times for the parameters $\removcoeff, \addcoeff = (-6, -3, \ldots, 3, 6)$, $\meandens = .25$.}, computing the current through the entire network at every step.  The results in \figref{fig:mega_picture} show that the system spontaneously forms channels (with high conductivity) at sufficiently negative $\removcoeff, \addcoeff$, and spontaneously forms walls (with low conductivity) at sufficiently positive $\removcoeff, \addcoeff$. This kind of phase transition is similar to those seen in driven lattice gas models \cite{Leung:1989vg, Schmittmann:1998cu}.

When the system channelizes due to negative \removcoeff, we find \emph{thin} channels, with glassy behavior; a negative \addcoeff gives \emph{thick} channels. To understand these transitions, we consider their robustness to perturbations. When the system has formed a set of parallel channels, occasionally a channel gets blocked (\figref{fig:NonEqAndFluctCartoon} c). When the channel is thin, current must go through the clog blocking the channel, and therefore total current through the channel is reduced while the clogging particle has much of this current forced through it. On the other hand, which this channel is thick, current will go around the clog, and so is barely impeded. Therefore, at negative $\addcoeff$, the thin channel has reduced current, and  this clogging will cause the thin channel to fill, while a large channel is much more robust and will not be filled. On the other hand, for negative $\removcoeff$, the clog in the wide channel has little current through it, and lingers, allowing the wide channel to eventually be filled; the clog in the thin channel has high current forced through it and is quickly removed. However, the system can become stuck in a glassy state-thick channels can form at negative $\removcoeff$ can persist, and multiple thick channels may persist at negative $\addcoeff$. This can lead to large hysteretic effects which are especially strong at very negative $\addcoeff, \removcoeff$, when the system is \quotes{frozen} and fluctuations are suppressed.  
 
For positive \removcoeff, the system forms thick walls, while for positive \addcoeff, the system forms a network of thin walls, although there does not appear to be a phase transition in this regime. In this state, occasionally a hole will form in the wall (\figref{fig:NonEqAndFluctCartoon} d). If the wall is thin, current rushes through the hole, and very little current goes through the rest of the wall. When the wall is thick, current through the wall and hole is roughly unchanged. Therefore, at positive $\removcoeff$, the thin wall will quickly disintegrate as a result of the hole, while the thick wall will persist, and the hole will eventually get filled. On the other hand, at positive \addcoeff, the hole in the thin wall will quickly be filled, while the hole in the thick wall will persist while more holes are allowed to appear. Interestingly, when \removcoeff is positive and \addcoeff is negative, both channelization \emph{and} wall-building phase separations occur, as the system phase-separates into thick clumps. For some choices of parameters(positive \addcoeff in \figref{fig:mega_picture}) we see scale-free correlations, where the strength of a mode depends not on the magnitude of the wave-vector, but only its direction. As we will see, both these features follow from a continuum model. 

 \begin{figure}[H] 
            \includegraphics[width=0.49\textwidth]{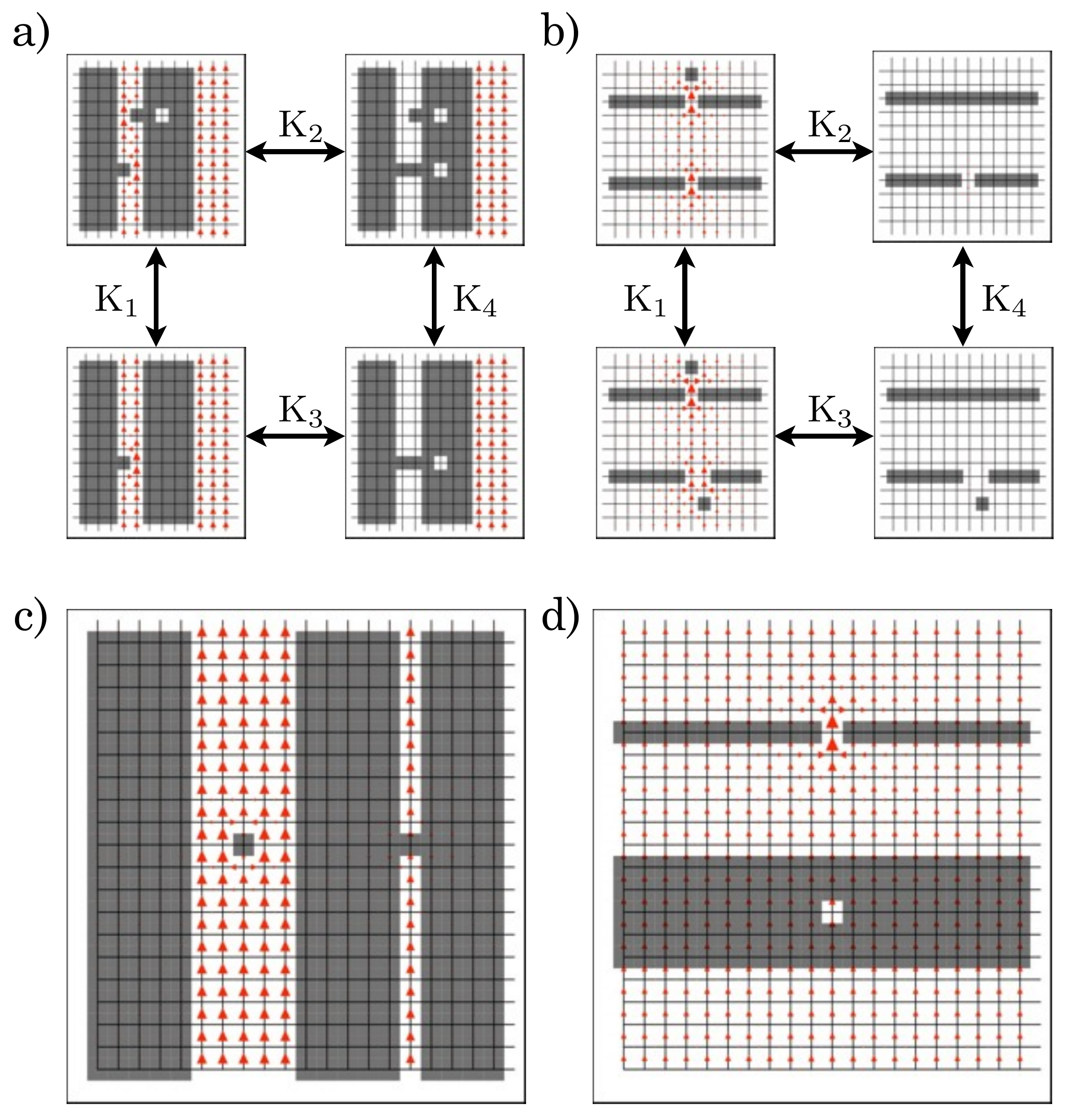}            
 \caption{ a), b): Lack of detailed balance. In a), $\addcoeff<0$, and $\text{K}_{2} \approx \text{K}_{3}$, as the current through the lower right vertex has only weak dependance on the occupation of the upper left vertex. However $\text{K}_{1} \not \approx \text{K}_{4}$, as the current through the upper left vertex strongly depends on the occupation of the lower right vertex; therefore $\text{K}_{1}  \text{K}_{2} \neq \text{K}_{3}  \text{K}_{4}$. A similar proof follows for situation b), where $\addcoeff >0$. c), d): Robustness of thick and thin structures(see text).}
 \label{fig:NonEqAndFluctCartoon}
\end{figure} 
To characterize the wall-building phase separation, we use the row density $\dens_{z} = \sum_{x} \dens_{xz}/\width$ as an order parameter, consistent with the fact that when the distribution of row densities becomes bimodal(\figref{fig:phase_sep_picture}d example),  wall-building has occurred. We characterize the channelization phase transition in terms of the column density  $\dens_{x} = \sum_{x} \dens_{xz}/\height$; when this is bimodal,  channelization has occurred (\figref{fig:mega_picture}). We characterize a clumping phase transition through local density $\bar{\dens}_{\posit} = 1/A  \ \sum_{\posit'} \theta(\sqrt{6} - \abs{\posit - \posit'}) \dens_{\posit'}$where $\theta$ is the Heaviside function. \footnote{we are limited to small system sizes, and thus a short-ranged local density function, for computational reasons. See \appendref{sec:bimodal} for details of determining bimodality.}; when this is bimodal and neither column or row density are bimodal, clumping has occurred.

\begin{figure}

   \includegraphics[width=0.4\textwidth]{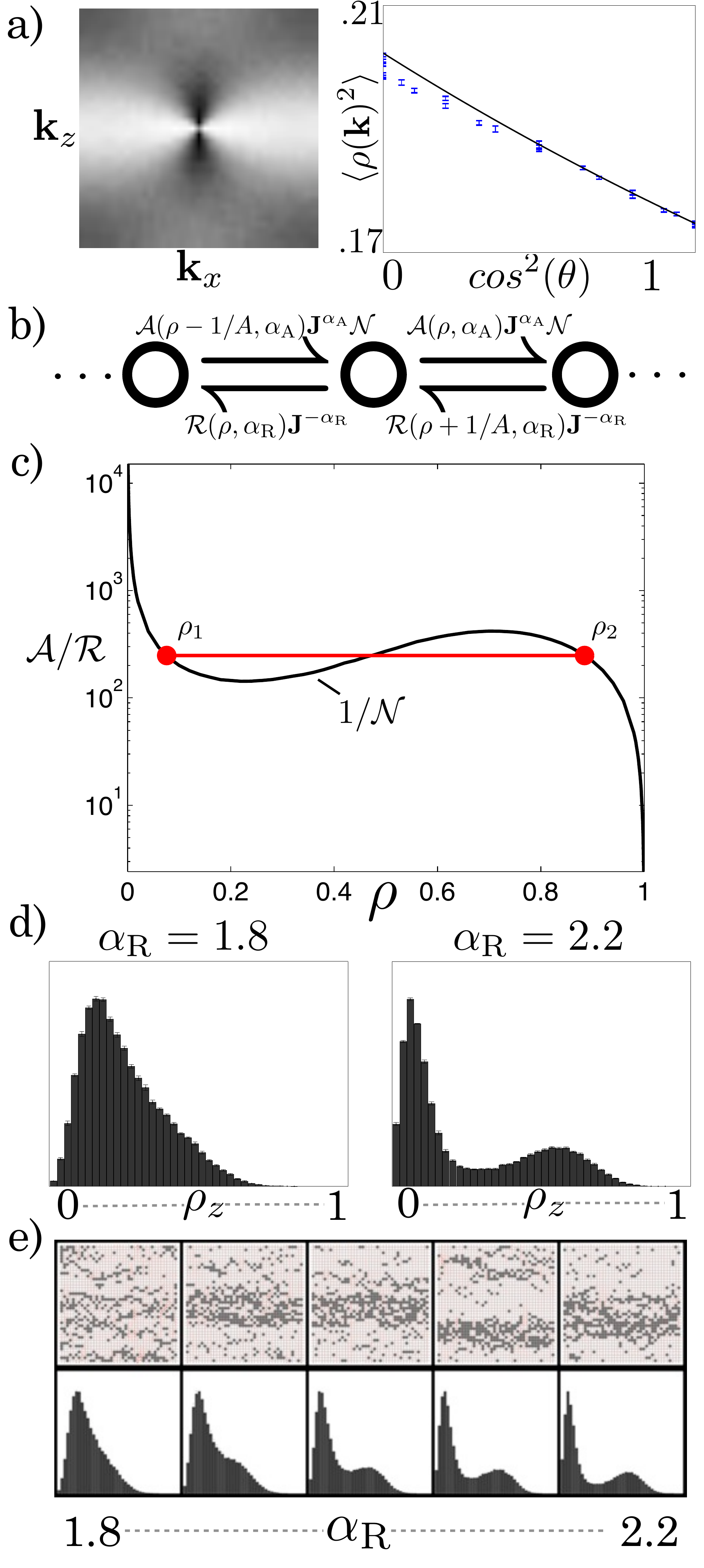} 
   \caption{
   a)$\expt{\dens(\wavevec)^{2}}$ for $(\removcoeff, \addcoeff) = (.1, .1)$. Scatter plot of $\expt{\dens(\wavevec)^{2}}$ vs. 
   $(\hat{\wavevec}\cdot \hat{x})^{2} = cos^{2}(\theta)$ for $(\abs{\wavevec}  \syssizesymbol)/2 \pi \leq 5$ compared with prediction from continuum model. Note that the majority of dependence is on  direction  $\wavevec_{x}^{2}/\abs{\wavevec^{2}}$, not magnitude $\abs{\wavevec}$.
    b) Fokker-Planck dynamics of continuum model for wall phase separation for a slice of area $A$. 
    c) Visualization of Eqs. \ref{eq:mass_bal}, \ref{eq:free_en}, for $(\removcoeff, \addcoeff) = (2.6, 0)$(color online)
    d) Comparison of histograms of row density, $\addcoeff = 0$ for a $\simsyssize \times \simsyssize$ system.  The right histogram is bimodal, so a wall phase transition is considered to have occurred. 
    e) Comparison of snapshots and row density histograms for wall-building phase transition.      
    }
   \label{fig:phase_sep_picture}
\end{figure}


\emph{Continuum model:}
Characterizing the distribution of filled vertices with a mean density field \dens, the continuum version of \myeqref{eq:discrete_circuit} is:
\myequation{
\meanflow =  \conduct(\dens)(- \nabla \voltage  + \grav \hat{z}) , \qquad
 \nabla \cdot \textbf{u} = 0
 \label{eq:darcys_law}
 }
where \conduct, the conductivity, is a function of density. Similarly, in the continuum limit, the discrete addition and removal activity are replaced with a stochastic equation for density evolution:
\myequation{
\dot{\dens} = - \meanremov(\dens, \meanflow, \removcoeff) +  \chempot\meanadd(\dens, \meanflow, \addcoeff) + \eta \label{eq:langevin}
}
where $\meanremov(\dens, \meanflow, \removcoeff)$ is the mean removal rate from a region of with density $\dens$, a current of $\meanflow$ and a bias of \removcoeff. \meanadd is the mean addition rate, and \meanflow is itself a functional of \dens obtained by solving \myeqref{eq:darcys_law}. $\chempot = \iint \meanremov/\iint \meanadd$ acts as a sort of chemical potential, which is set to conserve total particle number. $\eta$ is the stochastic noise term which creates fluctuations, whose form will be discussed later. \footnote{A complete continuum model should involve spatial terms to account for a short wavelength cutoff; here we do not include terms in $\meanadd, \ \meanremov, \ \conduct$ that involve spatial derivatives, focusing on the homogeneous mean field limit.}

The predictions of the continuum model depend strongly on the functions $\conduct, \meanadd, \meanremov$. To determine them, we use a \emph{hybrid} approach, sampling via numerical experiment using randomly placed particles and then varying the density to approximate the entire functions, leaving us with no fitting parameters \footnote{Making analytical approximations gives similar qualitative behavior, although the agreement with simulations is not as good (\appendref{app:MFTCartoon})} As in the discrete case, we start with a uniform density of $\meandens, \textbf{u}_{0} = \conduct(\meandens)$. \footnote{While our continuum model is a non-equilibrium system with no detailed balance, when certain limits and symmetries are assumed.}

\emph{Low \removcoeff, \addcoeff limit:}
In the limit where $\removcoeff, \addcoeff \to 0$, only the linear response is important. Writing \myeqref{eq:langevin} as
\myequationn{
\dot{\dens} = -\timederiv \paren{\dens, \meanflow, \vec{\alpha}} + \eta,
}
where $\timederiv\paren{\dens, \meanflow, \vec{\alpha}} = - \meanremov\paren{\dens, \meanflow, \removcoeff} + \chempot \meanadd \paren{\dens, \meanflow, \addcoeff}$ is the time derivative functional, $\vec{\alpha} = (\removcoeff, \addcoeff)$, we note that :
\myequationn{
\frac{d \timederiv}{d \dens}  =
\atfixed{\frac{\partial \timederiv}{\partial\dens}}_{\meanflow, \vec{\alpha}}   +
\atfixed{\frac{\partial \timederiv}{\partial \meanflow_{z}}}_{\dens, \vec{\alpha}} \frac{\partial \meanflow_{z}}{\partial \dens}.  \qquad
}
Furthermore, we decompose density fluctuations into its Fourier basis:
\myequationn{
\frac{d \timederiv(\wavevec)}{d \dens(\wavevec)}  =
\atfixed{\frac{\partial \timederiv}{\partial\dens}}_{\meanflow, \vec{\alpha}}   +
\atfixed{\frac{\partial \timederiv}{\partial \meanflow_{z}}}_{\dens, \vec{\alpha}} \frac{\partial \conduct}{\partial \dens}  \  \horizontality{\wavevec}, \qquad
}
where $\frac{\partial \meanflow_{z}(\wavevec)}{\partial \dens(\wavevec)} = \frac{\partial \conduct}{\partial \dens}  \  \horizontality{\wavevec}$ (see Supplementary Information). Characterizing $\eta$ as uncorrelated Gaussian noise via $\expt{\eta(\wavevec, t) \eta(\wavevec', t)} = 2 \delta(t - t') \delta_{\wavevec + \wavevec'} \noise,$
where $\noise = \bracket{ \meanremov(\meandens, \meanflow, \removcoeff)  + \chempot \meanadd(\meandens, \meanflow, \addcoeff) }/2$ is the effective diffusivity, we find that   $\dot{\dens}(\wavevec) = -\frac{d\timederiv(\wavevec)}{d \dens(\wavevec)}  \dens(\wavevec)+ \eta(\wavevec)$ to within first order. This allows us to predict the mean amplitude of fluctuations:
\myequationn{
\expt{\dens(\wavevec)^{2}} \approx \noise \paren{\atfixed{\frac{\partial \timederiv}{\partial\dens}}_{\meanflow, \vec{\alpha}}   +
\atfixed{\frac{\partial \meanflow_{z}}{\partial\dens}}_{\dens, \vec{\alpha}} \frac{\partial \conduct}{\partial \dens} \    \horizontality{\wavevec}}^{-1}
}
We note that this is independent of the magnitude of \wavevec, and is a function of its direction alone(\figref{fig:phase_sep_picture} a), because of the dipole-like interactions between particles which inhibit current upstream and downstream, while increasing it laterally. 

\def \mywallfact{\ensuremath{\meanflow^{\paren{-\removcoeff - \addcoeff}}} \xspace}
\def \mywalladdfact{\ensuremath{\meanflow^{{\addcoeff}}} \xspace}
\def \mywallremovfact{\ensuremath{\meanflow^{{-\removcoeff }} \xspace}}
\def \myotherchempot{\ensuremath{\tilde{\chempot}} \xspace}

\emph{Wall Phase Separation}
If we assume translation symmetry in the $\hat{x}$ direction $\meanflow(x, z)$ is constant throughout the system. In the discrete model the current through any vertex is proportional to $\meanflow$, so that we may make the simplification $\meanremov(\dens, \meanflow, \removcoeff) \to \meanremov(\dens, \removcoeff) \meanflow^{-\removcoeff}, \meanadd(\dens, \meanflow, \addcoeff) \to \meanadd(\dens, \addcoeff) \meanflow^{\addcoeff}$ \footnote{This is because in any region $v \propto \meanflow$, and therefore $\removfunct(v) \propto v^{-\removcoeff} \propto \meanflow^{-\removcoeff}$, $\addfunct(v) \propto v^{\addcoeff} \propto \meanflow^{\addcoeff}$}. 

We can view a horizontal slice containing $A$ vertices as having uniform density, obeying the dynamics shown in \figref{fig:phase_sep_picture}, with a mean addition rate $\meanadd(\dens, \addcoeff) \mywalladdfact \chempot$, and a mean removal rate of $\meanremov(\dens, \removcoeff) \mywallremovfact$. The first criteria for a phase separation to occur is mass balance between two horizontal slices of densities $\dens_{1}$, $\dens_{2}$:
\myequation{ 
\chempot = 
\negthinspace \frac{\meanremov(\dens_{1}, \removcoeff)\mywallremovfact}{\meanadd(\dens_{1}, \addcoeff)\mywalladdfact} = \
\negthinspace
 \frac{\meanremov(\dens_{2},  \removcoeff)\mywallremovfact}{\meanadd(\dens_{2},  \addcoeff)\mywalladdfact} \negthinspace = \
  \negthinspace \myotherchempot  \frac{\mywallremovfact}{\mywalladdfact} \label{eq:mass_bal}
}
where we have defined $\myotherchempot$ in order to separate the dependence of $\meanflow$ and $\dens$. Assuming each slice is large($A \to \infty$), we may find a recursion relation for the equilibrium distribution of densities:
\myequationn{
\frac{P(\dens + 1/A)}{P(\dens)} \approx {\frac{\meanadd(\dens', \addcoeff)\myotherchempot }{ \meanremov(\dens', \removcoeff)}},
}
giving conditions for free energy balance:
\myequation{ 
\int_{\dens_{1}}^{\dens_{2}}{ {\ln} \bracket{\frac{\meanadd(\dens', \addcoeff)\myotherchempot }{ \meanremov(\dens', \removcoeff)}} d \dens'} = 0. \label{eq:free_en}
}
The continuum model predicts the system to form walls when $\meandens$ falls between a satisfying $\dens_{1}, \dens_{2}$. Note that the wall phase separation is independent of $\meanflow$.

\emph{Channelization Phase Separation}
If we assume translation symmetry in the $\hat{z}$ direction, the mean current does not have to be uniform;  $\meanflow(x, z) = \conduct(\dens(x)) \hat{z}$. 
As before, each \emph{vertical} slice will obey the dynamics in \figref{fig:phase_sep_picture}, except the mean removal rate is now $ \meanremov(\dens, \removcoeff) \conduct(\dens)^{-\removcoeff}$, while the mean addition rate becomes $\chempot \meanadd(\dens, \addcoeff) \conduct(\dens)^{\addcoeff}$. Following the same procedure(see appendix for details), the criteria for mass balance becomes:
\myequationn{
\chempot = \frac{\meanremov(\dens_{1}, \removcoeff) }{\meanadd(\dens_{1}, \addcoeff)} \frac{\conduct(\dens_{1})^{-\removcoeff}}{\conduct(\dens_{1})^{\addcoeff}}=
 \frac{\meanremov(\dens_{2},  \removcoeff)}{\meanadd(\dens_{2},  \addcoeff)} \frac{\conduct(\dens_{2})^{-\removcoeff}}{\conduct(\dens_{2})^{\addcoeff}}
}
and the condition for free energy balance becomes
\myequationn{
\int_{\dens_{1}}^{\dens_{2}}{ {\ln} \bracket{\frac{\meanadd(\dens', \addcoeff)\chempot \conduct(\dens')^{\addcoeff + \removcoeff} }{ \meanremov(\dens', \removcoeff) }} d \dens'} = 0. 
}
Note that when $\addcoeff = -\removcoeff$, the criteria for wall-building and channelization become identical; if a phase separation occurs, it will occur in both orientations, giving rise to a clumping phase transition. This is what we have observed in simulations. 

Our continuum model predicts the formation of walls, channels and clumps. However, the functions characterizing conductivity and activity used in this continuum model come from numerical experiments which neglect microscopic correlations, resulting in an incorrect prediction of the order of the phase transition. In addition, because there is no inherent length scale to the continuum model, it can not explain the transition between thin and thick structures.  A continuum model considering the formation of the thinnest structures also predicts channelization and walling (\appendref{app:crystalMFT}), but a theory combining both elements has no additional predictive power. Additionally the model predicts scale-free dipole-like correlations  observed in the discrete model, which ought to exist in all nearly-disordered systems with these properties.

\emph{Discussion:}
Our model relies on very simple elements found across multiple living and nonliving systems; indeed, a coarse-grained view would often yield the same model of a stochastically evolving porous medium where the resistance and flow are coupled to each other, in the presence of a conserved current. Despite this simplicity, our discrete numerical simulations show channeling and walling phase separations at multiple length scales and orientations consistent with the biases of systems it is inspired by. However, whether or not the entire phase diagram of possible configurations is explored in natural systems remains unknown.

Acknowledgments: We thank Mehran Kardar for discussions about the nature and order of the phase transitions, and suggesting the short length scale mean-field theory. For partial financial support, we thank the Henry W. Kendall physics fellowship (S.O), the Wyss Institute and the MacArthur Foundation (L.M) and Human Frontiers Science Program grant RGP0066/2012-TURNER(S.O., L.M.).  

\bibliography{APMBib}

\clearpage
\pagebreak

\onecolumngrid
\linespread{1.} 

\appendix

\section{Determining if a histogram is bimodal}\label{sec:bimodal}

Each simulation $i$ at a particular parameter value gives us a $P_{i}(N)$, the probability of measuring a $N$ particles in a row, column, or clump in simulation. Averaging many individual simulations allows us to calculate an average $P(N)$, as well as $\sigma(N)$, the estimated standard deviation of this average measurement. We use a 3-sigma threshold of statistical significance-we are significantly more likely to measure $N$ particles than $N'$ iff 

\myequation{
P(N) - P(N') > 3 \sqrt{\sigma(N)^{2} + \sigma(N')^{2}}
\label{app:statsignif}
}

 If a local maximum $P(N)$ can reach a larger $P(N'')$ without moving through a valley where \eqref{app:statsignif} holds, it is considered to be a \emph{false peak}. If not, it is considered to be a \emph{true peak}. A histogram with at least two true peaks is considered to be bimodal.

\begin{figure*}[h!]
   \includegraphics[width=.6\textwidth]{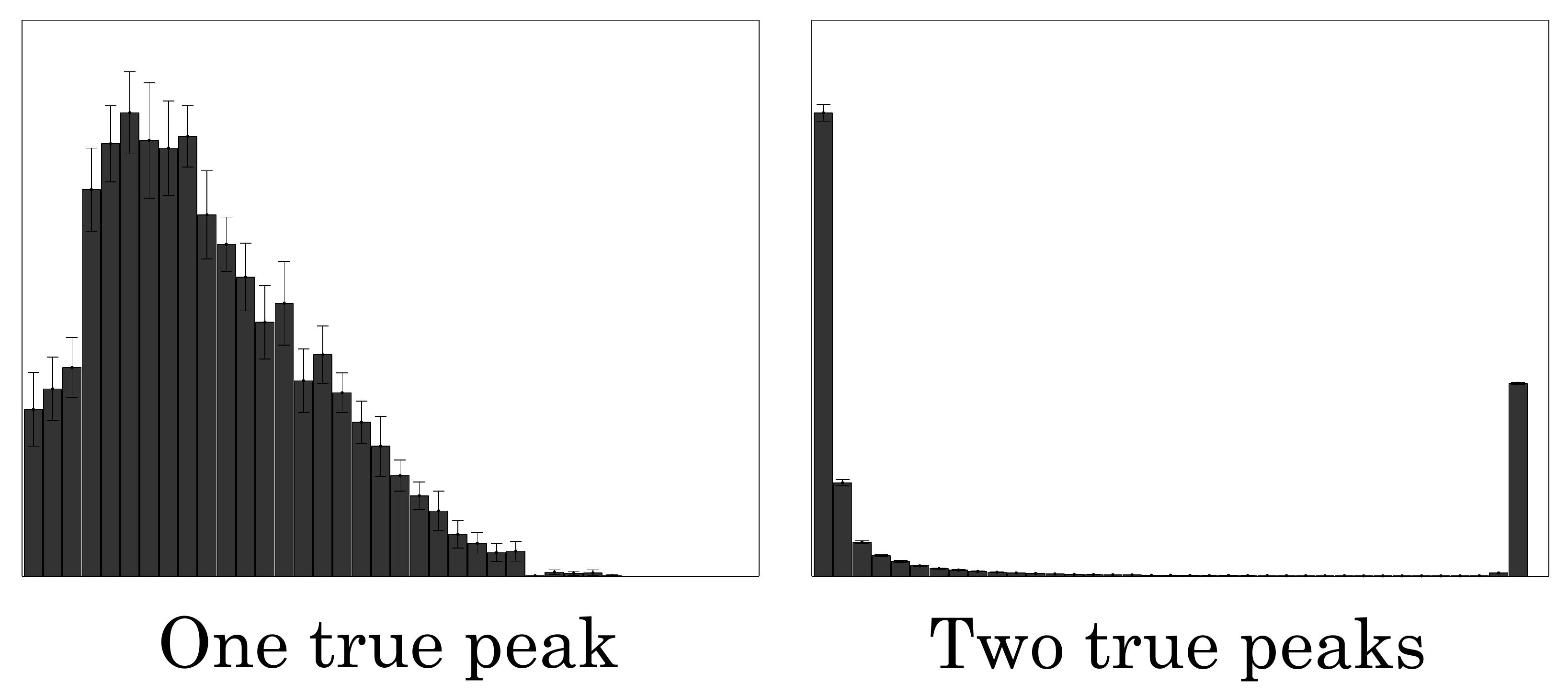} 
   \caption{The distribution on the left has many false peaks, but is not considered to be bimodal. The right distribution is.}
   \label{fig:skew_picture}
\end{figure*}

\section{Comparison to local dynamics}\label{app:LocalDynamics}
For local dynamics, we we use a slightly modified time integration step. 

\vspace{-2 mm}
\begin{enumerate}\itemsep0pt \parskip 0 pt \parsep 0 pt
\item Remove a particle from filled vertex  $i$ randomly selected with probability proportional to $\removfunct(v_{i})$.
\item Solve for the new current through the network.
\item Add the particle to an empty vertex $j$ randomly selected with probability proportional to ${\addfunct(v_{j}) \ e^{-\frac{ (\posit_{j} - \posit_{i})^{2}}{2 \diffus^{2}}}}$.
\item Solve for the new current through the network.
\end{enumerate}
\vspace{-2 mm}

This prohibits a removed particle from traveling non-locally. 

\begin{figure*}[h!]
   \includegraphics[width=.8\textwidth]{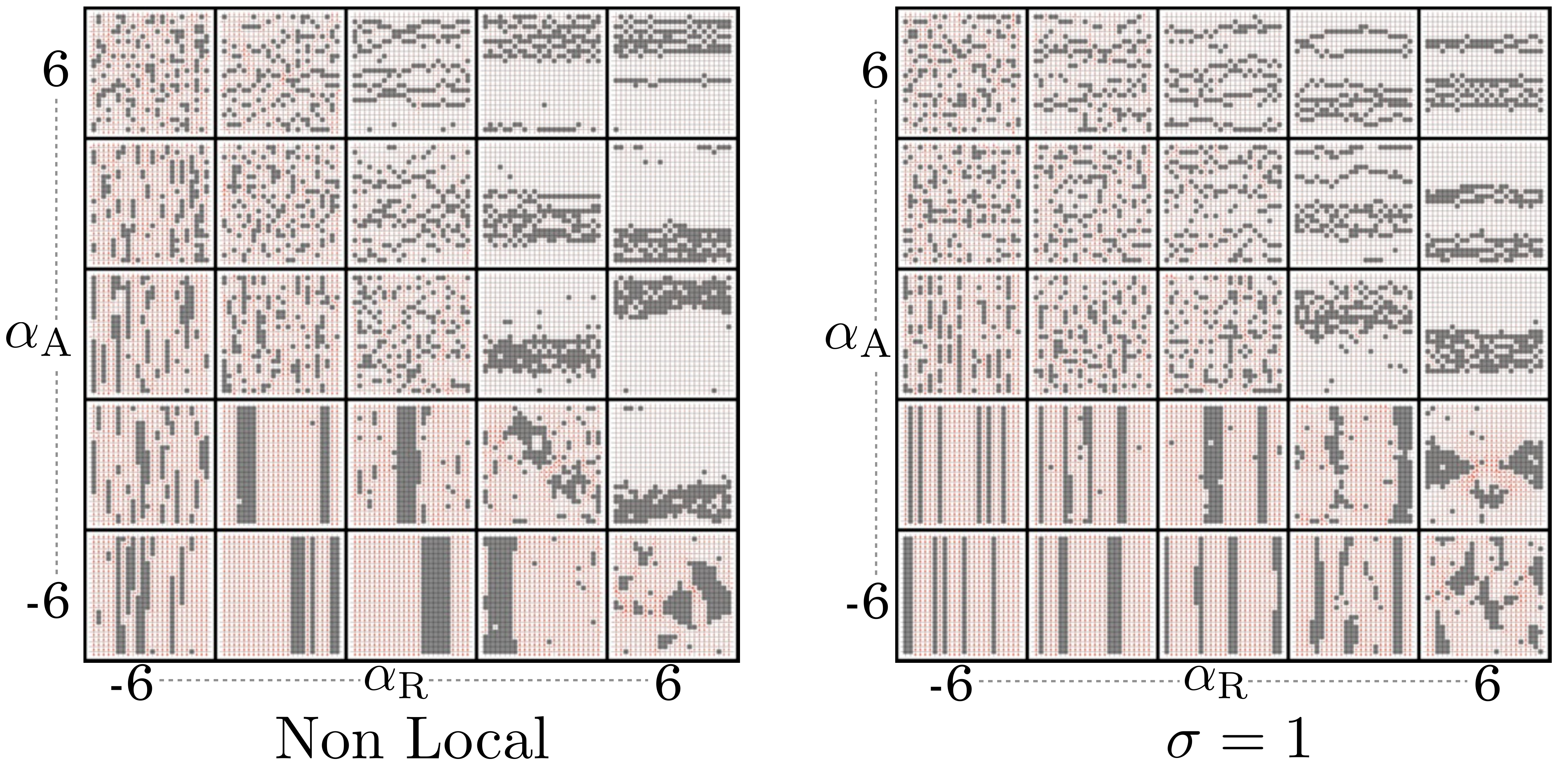} 
   \caption{Comparison of Nonlocal and Local Dynamics}
   \label{fig:NonLocalVsLocal}
\end{figure*}

The phase diagram generated is very similar (\figref{fig:NonLocalVsLocal}).

\section{Analytical Mean Field Theory}\label{app:MFTCartoon}
An alternate mean field theory produces some of the same qualitative behavior. Instead of relying on numerics to find the values of $\meanadd(\dens, \addcoeff), \ \meanremov(\dens, \removcoeff), \ \conduct(\dens)$, we rely on a very simple model which gives analytical results. 

\begin{figure*}[h!]
   \includegraphics[width=.3\textwidth]{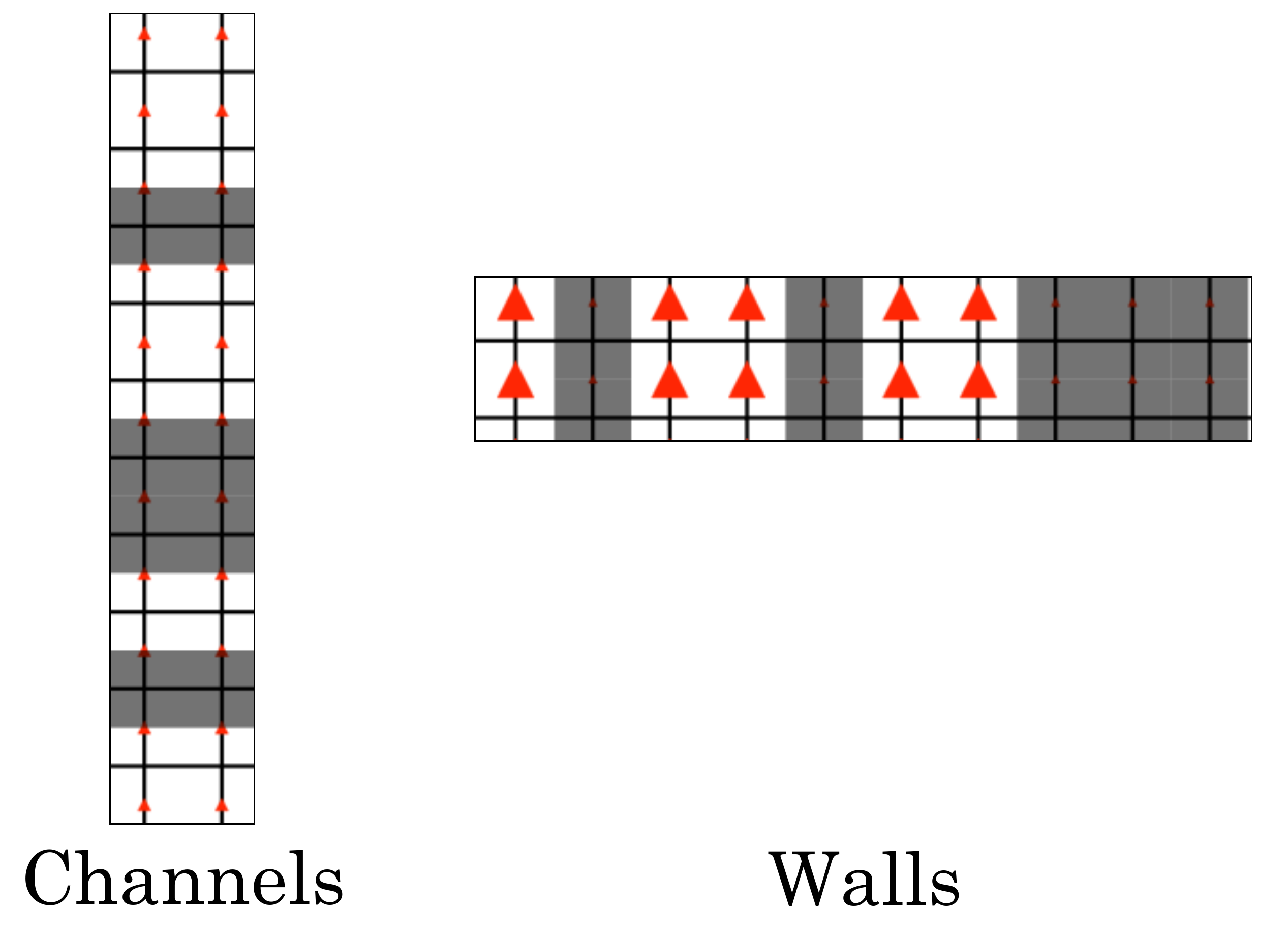} 
   \caption{Schematic of approximations made to obtain analytic forms for $\meanadd, \meanremov, \conduct$ for channelization and wall-building}
   \label{fig:dummy_MFT_example}
\end{figure*}

\emph{Channels}:
For predicting a channelization phase transition, we assume that current is unable to travel in the $\hat{x}$ direction(\figref{fig:dummy_MFT_example}). Therefore, the equation for conductivity is:

\def \chan{\text{chan}}
\def \wall{\text{wall}}

\myequationn{
\conduct_{\chan}(\dens) = \frac{1}{1 + \Delta \resist \dens}.
}
Because current cannot flow laterally, an equal current of $\meanflow$ is pushed through the filled and empty vertices, and so 

\myequationn{
\meanadd_{\chan}(\dens, \addcoeff)  = (1-\dens), \qquad
\meanremov_{\chan}(\dens, \removcoeff) = \dens.
}

We note that $\frac{\meanadd_{\wall} \conduct_{\wall}^{\addcoeff}}{\meanremov_{\wall} \conduct_{\wall}^{-\removcoeff}}$ is a function of $\addcoeff + \removcoeff$, and has no individual dependence on $\addcoeff, \removcoeff$.

\emph{Walls:}
For predicting a wall phase transition, we assume that, between rows, current can freely flow in the $\hat{x}$ direction without any resistance (\figref{fig:dummy_MFT_example}). Therefore, the equation for conductivity is 

\myequationn{
\conduct_{\wall}(\dens) = (1- \dens) + \frac{\dens}{1 + \Delta \resist}.
}

The total driving across a wall is $\frac{\meanflow}{\conduct_{\wall}}$, and thus the current across an empty vertex is $\frac{\meanflow}{\conduct_{\wall}}$.  The total current across a filled vertex is $\frac{\meanflow}{\conduct_{\wall}(1 + \Delta \resist)}$. Therefore:

\myequationn{
\meanadd_{\wall}(\dens, \addcoeff)  = (1-\dens) \paren{\frac{1}{\conduct_{\wall} (\dens)  }}^{-\removcoeff}, \qquad
\meanremov_{\wall}(\dens, \removcoeff) = \dens \paren{\frac{1}{\conduct_{\wall}(\dens) \ (1 + \Delta \resist)}}^{\addcoeff}. 
}
We note that $\frac{\meanadd_{\wall}}{\meanremov_{\wall}}$ is also function of $\addcoeff + \removcoeff$, and has no individual dependence on $\addcoeff, \removcoeff$.\\

When $\meandens = .25$, a channelization phase transition occurs when $\addcoeff + \removcoeff \lesssim -1.55$. A wall-building phase transition occurs when $\addcoeff + \removcoeff \gtrsim 2.25$.


\def \skewcoeff{\ensuremath{\mathfrak{s}}\xspace}
\def \regfrac{\ensuremath{\mathcal{V}}\xspace}
\def \myratio{\ensuremath{\mathcal{N}_{\text{}}}\xspace}

\section{Short Length Scale Continuum Model}\label{app:crystalMFT}
We can also create a continuum model on a short length scale. To do so, we select a periodic structure with two regions labeled 1 and 2. Region 1 comprises a fraction $\regfrac_{1}$ of the squares, while region 2 comprises a fraction $\regfrac_{2}$ of squares.

The density will originally be uniform, s.t. $\meandens_{1} = \meandens_{2} = \meandens$, and a mean density of \skewcoeff (\figref{fig:skew_example}) can transfer between squares such that

\myequationn{
\meandens_{1} = \meandens + \skewcoeff/\regfrac_{1},  \qquad \meandens_{2} = \meandens - \skewcoeff/\regfrac_{2}
}

\begin{figure*}[h!]
   \includegraphics[width=.7\textwidth]{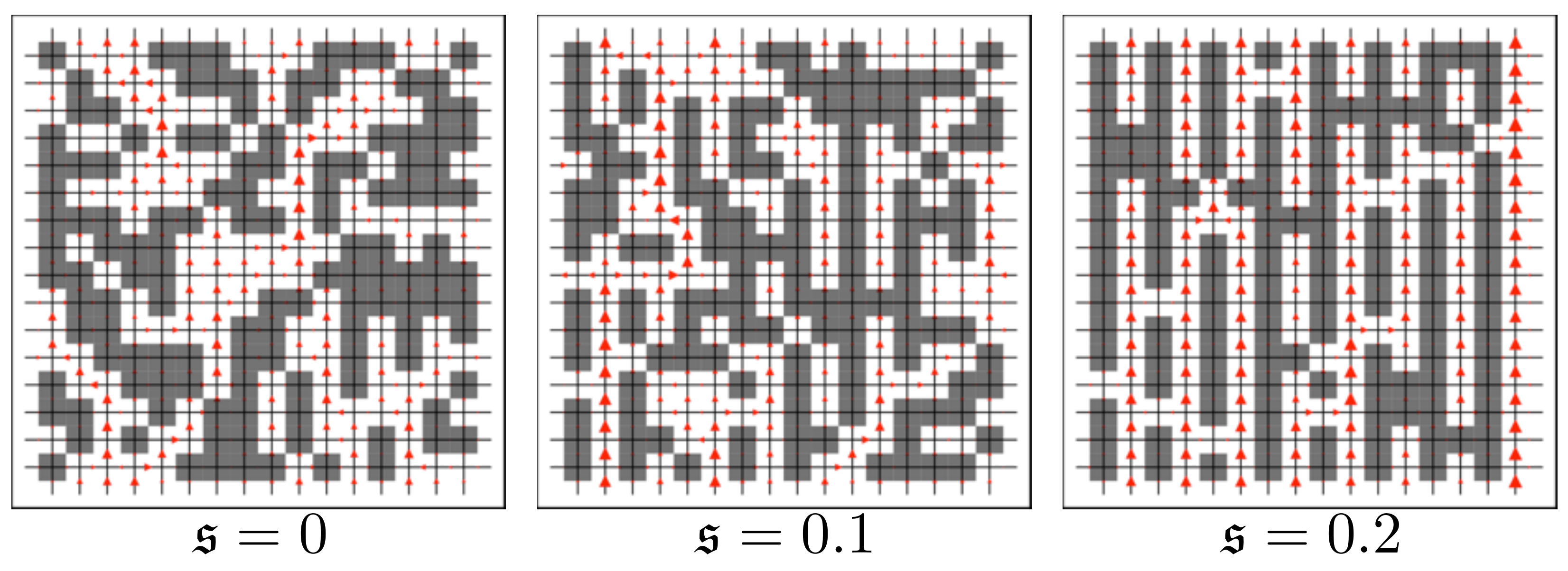} 
   \caption{ Example systems where $\meandens = .5$, $\skewcoeff = 0, 0.1, 0.2$, with a spacing of $d = 2$. Orientation is set to channels/pillars. }
   \label{fig:skew_example}
\end{figure*}

At a particular imbalance $\skewcoeff$, the probability of an particle moving from region 2 to region 1 divided by the probability of the opposite process gives us a \quotes{fugacity}:

\myequationn{ \myratio(\meandens, \skewcoeff, \addcoeff, \removcoeff) = 
	\frac{\meanremov_{2} \paren{\meandens_{2}, \conduct(\meandens, \skewcoeff), \removcoeff} \cdot \meanadd_{1} \paren{\meandens_{1}, \conduct(\meandens, \skewcoeff), \addcoeff} }	
	{\meanremov_{1} \paren{\meandens_{1}, \conduct(\meandens, \skewcoeff), \removcoeff}  \cdot \meanadd_{2} \paren{\meandens_{2}, \conduct(\meandens, \skewcoeff), \addcoeff}}
}

Therefore, the free energy of a state with an imbalance \skewcoeff is 
\myequationn{
-\int_{0}^{\skewcoeff} \text{Ln} \bracket{\myratio(\meandens, \skewcoeff', \addcoeff, \removcoeff)} d \skewcoeff'
} 
\skewcoeff will be set to minimize free energy, and when the optimal \skewcoeff is nonzero, the continuum model predicts the system to spontaneously \quotes{crystallize} into a form where regions 1 and 2 have different density.
For thin channels, region 1 is set by $\delta_{x \text{ mod } d}$, where d some integer which sets the spacing between channels or pillars. This walls are the same except that region 1 is now set by $\delta_{y \text{ mod } d}$.

The short-length scale continuum model gives similar behavior to the uniform continuum model, although the change in free energy is nearly always lower.

\begin{figure*}[h!]
   \includegraphics[width=.8\textwidth]{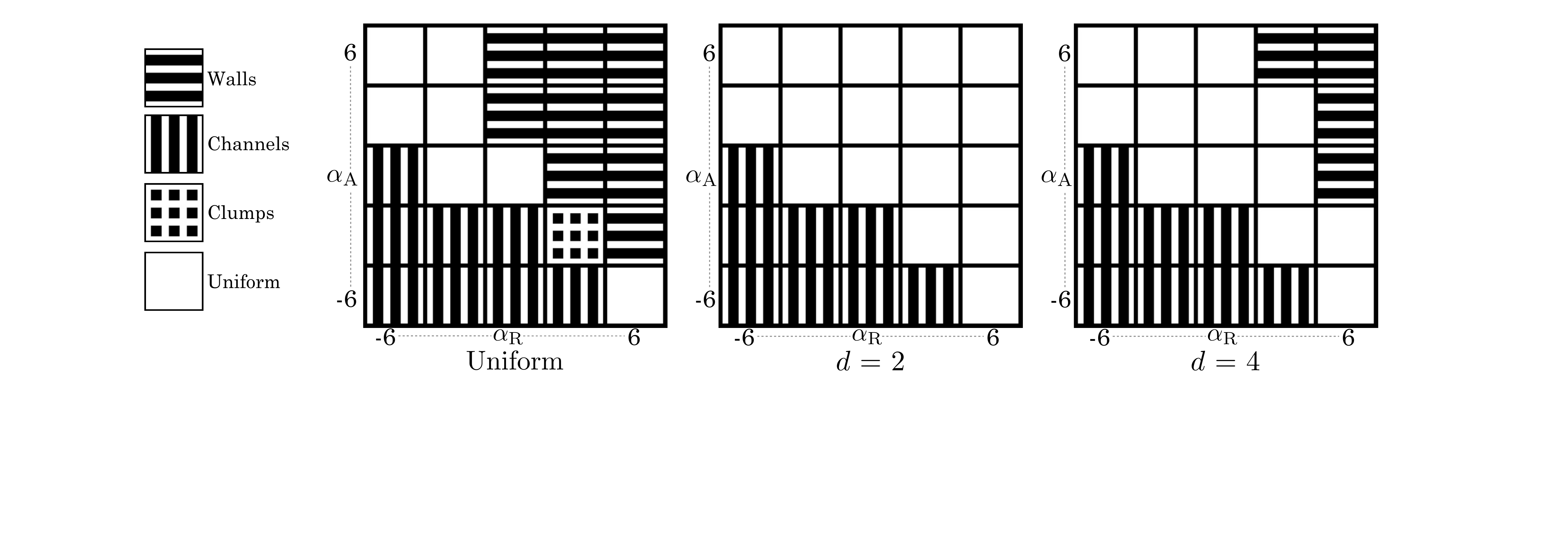} 
   \caption{Comparison of large and small length scale continuum model where $\meandens = .25$}
   \label{fig:skew_picture}
\end{figure*}

\newcommand{\myphase}[1]{e^{i{#1} \cdot \posit}}
\newcommand{\decomposedvec}[1]{#1_{x} \hat{x} + #1_{z} \hat{z}}
\newcommand{\undecomposedvec}[1]{#1}
\newcommand{\xcomp}[1]{#1_{x}}
\newcommand{\zcomp}[1]{#1_{z}}

\def \mysecorddelt{\delta_{\wavevec' + \wavevec'', \wavevec} \xspace}
\def \condcond{\conduct_{1}(\wavevec') \conduct_{1}(\wavevec'') \xspace}
\def \densdens{\dens(\wavevec') \dens(\wavevec'') \xspace}
\def \condconddelt{\conduct_{1}(\wavevec') \conduct_{1}(\wavevec'')\mysecorddelt \xspace}
\def \densdensdelt{\dens(\wavevec') \dens(\wavevec'')\mysecorddelt \xspace}

\section{Low $\alpha$ limit}
We start off from a Langevin equation, 
\myequation{
\dot{\dens} = -\timederiv \paren{\dens, \meanflow(\dens), \vec{\alpha}} + \eta, \label{eq:applangevin}
}
Where \timederiv is the time derivative functional, $\vec{\alpha}$ is short for $  (\removcoeff, \addcoeff)$ =  and \meanflow is itself a functional of \dens obtained by solving \eqref{eq:darcys_law}. We decompose this:
\myequationn{
\frac{d \timederiv}{d \dens}  =
\atfixed{\frac{\partial \timederiv}{\partial\dens}}_{\meanflow, \vec{\alpha}}   +
\atfixed{\frac{\partial \timederiv}{\partial \meanflow_{z}}}_{\dens, \vec{\alpha}} \frac{\partial \meanflow_{z}}{\partial \dens}  \qquad
\atfixed{\frac{\partial \timederiv}{\partial \meanflow_{x}}}_{\dens, \vec{\alpha}} \frac{\partial \meanflow_{x}}{\partial \dens} 
}
We note that, due to symmetry, the third term is zero and may be removed. Moving into fourier space, where $\dens = \dens_{0} + \iint \dens(\wavevec) \myphase{\wavevec}$, we we have: 
\myequationn{
\frac{d \timederiv(\wavevec)}{d \dens(\wavevec)}  =
\atfixed{\frac{\partial \timederiv}{\partial\dens}}_{\meanflow, \vec{\alpha}}   +
\atfixed{\frac{\partial \meanflow_{z}(\wavevec)}{\partial\dens(\wavevec)}}_{\dens, \vec{\alpha}} \frac{\partial \meanflow_{z}}{\partial \dens}.
}
We now must find:
\myequationn{
\frac{d\meanflow_{z}(\wavevec)}{d \dens(\wavevec)}. 
}
To do so, we start off with a uniform density $\dens_{0}$ and then apply a sinusoidal perturbation $\Delta \dens \myphase{\wavevec}$. The conductivity is, to within first order

\myequationn{
\conduct = \conduct_{0} + \frac{d\conduct}{d \dens} \Delta \dens \myphase{\wavevec} = \conduct_{0} + \Delta \conduct \myphase{\wavevec}
}
Giving us a mean current
\myequationn{
\meanflow = \bracket{\conduct_{0} + \Delta \conduct \myphase{\wavevec}} \hat{z}  - \conduct_{0} \nabla \bracket{ \Delta \voltage \myphase{\wavevec}} = \conduct_{0} \hat{z} + \Delta \conduct \hat{z} = \meanflow_{0} + \Delta \meanflow \myphase{\wavevec}, \ \ 
\Delta \meanflow =  \myphase{\wavevec} \bracket{\Delta \conduct  \hat{z}  - \conduct \Delta \voltage \bracket{i k_{x} \hat{x} + i k_{z} \hat{z}  } }
}
We set $\Delta \voltage$ to conserve current up to first order:
\myequationn{
\nabla \cdot \Delta \meanflow = 
   \myphase{\wavevec} \bracket{ \Delta \conduct i k_{z} + \conduct_{0} \Delta \voltage \bracket{k_{x}^{2} + k_{z}^{2}}} = 0 
\Rightarrow \Delta \voltage =   \frac{-i k_{z}}{\abs{k^{2}}\conduct_{0}}. 
}
Giving us our change in current,
\myequation{
\Delta \meanflow 
= \Delta \dens \frac{d \conduct}{d \dens} \bracket{\hat{z} + \frac{ik_{z}}{|k|^{2} }\bracket{i k_{x} \hat{x} + i k_{z} \hat {z}} } 
= \Delta \dens \frac{d \conduct}{d \dens}  \bracket{ \frac{k_{x}^{2}}{|k|^{2}} \hat{z} -  \frac{k_{x}k_{y}}{|k|^{2}} \hat{x}}
\Rightarrow
\frac{d\meanflow_{z}(\wavevec)}{d \dens(\wavevec)} = \frac{d \conduct}{d \dens} \horizontality{\wavevec}, 
\label{eq:firstordercurrentderiv}
}
Plugging \ref{eq:firstordercurrentderiv} into \ref{eq:applangevin} yields:
\myequationn{
	\dot{\dens}(\wavevec) = -\paren{\atfixed{\frac{\partial \timederiv}{\partial\dens}}_{\meanflow, \vec{\alpha}}   +
\atfixed{\frac{\partial \meanflow_{z}}{\partial\dens}}_{\dens, \vec{\alpha}} \frac{\partial \conduct}{\partial \dens}    \horizontality{\wavevec}} \dens(\wavevec)  + \eta(\wavevec)
}
where $\expt{\eta(\wavevec, t) \eta^{*}(\wavevec, t') } = 2 \delta(t - t') \noise$. The Einstein relation then predicts the strength of fluctuations to within first order:
\myequationn{
\expt{\dens(\wavevec)^{2}} \approx \noise \paren{\atfixed{\frac{\partial \timederiv}{\partial\dens}}_{\meanflow, \vec{\alpha}}   +
\atfixed{\frac{\partial \meanflow_{z}}{\partial\dens}}_{\dens, \vec{\alpha}} \frac{\partial \conduct}{\partial \dens} \cdot   \horizontality{\wavevec}}^{-1} .
}

\newcommand{\Deltadenscomp}[1]{\Delta \dens \paren{{#1}}\text{ }\myphase{#1} }

\def \bigO{\mathcal{O}}
\def \thirdordstuff{\bigO(\Delta \dens^{3}) \xspace}
\def \finalinteraction{F \xspace}
\section{Higher Order Terms}

To find higher order dependencies of $\timederiv$ of $\Delta \dens$, we go through the following steps:
\begin{enumerate}
\item Obtain $\Delta \conduct$ from $\Delta \dens$ up to the desired order. 
\item Obtain $\voltage$ setting flow to be conserved up to the desired order, e.g. ${
\nabla \cdot \meanflow = \nabla \cdot \bracket{\conduct \hat{z} -\conduct \nabla \voltage}= 0
}$
\item Obtain $\meanflow$ through $\meanflow = \conduct \bracket{\hat{z} - \nabla \voltage}$
\item Obtain $\timederiv$ through \dens, \meanflow up to the desired order. 
\end{enumerate}

Here, we will carry out all steps to within second order. We do not go higher as the number of terms increases quite rapidly.\\

\subsection{Finding change in conductivity}
 We start from 
\myequationn{
\dens = \dens_{0} + \Delta \dens =  \dens_{0} + \sum_{\wavevec}  \dens(\wavevec) \myphase{\wavevec}
}
First we must find the change in conductivity, which we split into first and second order components: $\conduct = \conduct_{0} + \conduct_{1} + \conduct_{2} + \bigO(\Delta \dens^{3})$
\myequationn{
\conduct = \conduct_{0}  + \Delta \conduct = \conduct_{0} + \conduct_{\dens} \Delta{\dens} + \frac{1}{2} \conduct_{\dens \dens} \Delta \dens ^{2} + \thirdordstuff =\\
 \conduct_{0} + \conduct_{\dens} \sum_{k}\Deltadenscomp{k} + \frac{1}{2} \conduct_{\dens \dens} \sum_{k', k''}  \dens(\wavevec') \dens(\wavevec'') \myphase{(\wavevec' +\wavevec'')} + \thirdordstuff.
} 

Giving us the first order and second order change in conductivity $\conduct_{1}, \conduct_{2}$:
\myequation{
\conduct_{1}(\wavevec) = \conduct_{\dens}\dens(\wavevec) \label{eq:firstordercond}  \\ 
\conduct_{2}(\wavevec) = \frac{1}{2} \conduct_{\dens \dens} \dens(\wavevec')\dens(\wavevec'')\mysecorddelt\label{eq:secordercond}.
}

\subsection{Change in voltage} 
We must now find the change in voltage, which we do by setting current to be conserved. 
\myequationn{
\nabla \cdot \meanflow = \nabla \cdot \bracket{\conduct \hat{z} -\conduct \nabla \voltage} = \conduct_{z} - \conduct \nabla^{2} \voltage - \nabla \conduct \cdot \nabla\voltage +\thirdordstuff= 0
}

We separate voltage into it's first and second order components $\voltage_{1}, \voltage_{2}$:

\newcommand{\voltageone}[1]{\ensuremath{
\frac{-i  #1_{z}}{\conduct_{0}\abs{#1}^{2}}\conduct_{1}(#1)
}
}

\myequationn{
\partial_{z}\conduct_{1} - \conduct_{0} \nabla^{2} \voltage_{1} + \partial_{z}\conduct_{2}- {\conduct_{1}}\nabla^{2}\voltage_{1} - \conduct_{0} \nabla^{2} \voltage_{2} - \nabla \conduct_{1} \cdot \nabla \voltage_{1} = 0 
}
We solve for $\voltage_{1}$ by balancing all first-order terms:
\myequation{
\nabla^{2} \voltage_{1} = \sum_{\wavevec} -\abs{\wavevec}^{2} \voltage_{1}(\wavevec) \myphase{\wavevec}= 
 \frac{\partial_{z}\conduct_{1} }{\conduct_{0}} = \frac{1}{\conduct_{0}}\sum_{\wavevec} i k_{z} \conduct_{1} \paren{\wavevec} \ \myphase{\wavevec} 
 \\ \Rightarrow
 \voltage_{1}(\wavevec) = \frac{-i k_{z}}{\conduct_{0}\abs{\wavevec}^{2}}  \conduct_{1}(\wavevec). \label{eq:firstordervolt}
%
}
To calculate voltage to second order, there are four contributions which must be matched;
\myequationn{
\partial_{z}\conduct_{2} - \conduct_{1} \nabla^{2} \voltage_{1} - \conduct_{0} \nabla^{2} \voltage_{2} - \nabla \conduct_{1} \cdot \nabla \voltage_{1} = 0
}
Fortunately, there is only one factor of $\voltage_{2}$ which we can move to the left hand side:
\myequationn{
\conduct_{0} \nabla^{2} \voltage_{2}  = \partial_{z}\conduct_{2} -\conduct_{1} \nabla^{2} \voltage_{1} - \nabla \conduct_{1} \cdot \nabla \voltage_{1}\\
\Rightarrow
\sum_{\wavevec}{ \paren{-\conduct_{0} \abs{\wavevec}^{2}} \voltage_{2}(\wavevec) \myphase{\wavevec} } = 
\underbrace{\sum_{\wavevec}{ i \wavevec_{z} \conduct_{2}(\wavevec)\myphase{\wavevec} } }_{\partial_{z}\conduct_{2} }  
 - \sum_{\wavevec' \wavevec''}{ \voltage_{1}(\wavevec')\conduct_{1}(\wavevec'')
\bracket{
\underbrace{-\abs{\wavevec'}^{2}}_{\conduct_{1} \nabla^{2} \voltage_{1}} -
\underbrace{ \wavevec' \cdot \wavevec''}_{\nabla \conduct_{1} \cdot \nabla \voltage_{1}}
}
\myphase{(\wavevec' + \wavevec'')} }
}
\def \almostfinalvolttwo{
 \paren{\conduct_{0} \abs{\wavevec}^{2}} ^{-1} 
\bracket{
-i \wavevec_{z} \conduct_{2}(\wavevec)+ 
{ 
{\frac{i \wavevec'_{z}}{\conduct_{0}}} 
\paren{
{ \frac{\abs{\wavevec'}^{2}  +{\wavevec' \cdot \wavevec''} }{\abs{\wavevec'}^{2} }}
}
  \conduct_{1}(\wavevec')\conduct_{1}(\wavevec'')  \mysecorddelt} 
}
}

\def \finalvolttwo{
 \paren{\conduct_{0} \abs{\wavevec}^{2}} ^{-1} 
\bracket{
-i \wavevec_{z} \conduct_{2}(\wavevec)+ 
{ 
{\frac{i \wavevec'_{z}}{\conduct_{0}}} 
\paren{
{ \frac{\wavevec' \cdot \wavevec }{\abs{\wavevec'}^{2} }}
}
  \conduct_{1}(\wavevec')\conduct_{1}(\wavevec'')  \mysecorddelt} 
}
}

\myequationn{
\Rightarrow \voltage_{2}(\wavevec) =  \almostfinalvolttwo 
}

Which we then simplify using $\wavevec  \ \mysecorddelt= \paren{\wavevec' + \wavevec''} \mysecorddelt$:

\myequation{
\Rightarrow \voltage_{2}(\wavevec) =  \finalvolttwo \label{eq:secordervolt}.
}

\subsection{Change in mean flow}
We start off from the equation for $\meanflow$:

\myequationn{
\meanflow = \conduct \paren{\hat{z} - \nabla \voltage}
}
Accumulating first and second order terms
\myequationn{
\meanflow = \conduct_{0} \hat{z} + \conduct_{1} \hat{z}  + \conduct_{2} \hat{z}- \conduct_{0} \nabla \voltage_{1} - \conduct_{0} \nabla \voltage_{2} - \conduct_{1} \nabla \voltage_{1}  + \thirdordstuff
}
We first calculate the first-order current, $\meanflow_{1}$.
\myequationn{
\meanflow_{1} = \conduct_{1} \hat{z} - \conduct_{0}  \nabla \voltage_{1} = 
\sum_{\wavevec} \conduct_{1}(\wavevec) \myphase{\wavevec} \hat{z} - 
\conduct_{0} \sum_{\wavevec} \conduct_{1} (\wavevec) i \paren{\decomposedvec{\wavevec}} \frac{-i k_{z}}{\conduct_{0}\abs{\wavevec}^{2}} \wavevec_{z}   = 
\sum_{\wavevec} \conduct_{1}(\wavevec) \myphase{\wavevec} 
\bracket{\hat{z} {\frac{\wavevec_{x}^{2}}{\abs{\wavevec}^{2} }}    - \hat{x} \frac{\wavevec_{x}\wavevec_{z}}{\abs{\wavevec}^{2}}    }
}
\myequation{ 
\Rightarrow
\meanflow_{1}(\wavevec)  = \conduct_{1}(\wavevec) \bracket{\hat{z}   \horizontality{\wavevec}  - \hat{x} \frac{\wavevec_{x}\wavevec_{z}}{\abs{\wavevec}^{2}}    } 
\label{eq:firstorderflow}.
}

Calculating the second-order current $\meanflow_{2}$, we find:
\myequationn{
\meanflow_{2} =  
\sum_{\wavevec} \meanflow_{2}(\wavevec) \myphase{\wavevec} =\conduct_{2} \hat{z} - \conduct_{1} \nabla \voltage_{1} - \conduct_{0} \nabla \voltage_{2} 
}
We expand this as:
\myequationn{
\sum_{\wavevec} \meanflow_{2}(\wavevec) \myphase{\wavevec} 
 = \sum_{\wavevec} \conduct_{2} \hat{z}  -
 \sum_{\wavevec', \wavevec''} \voltage_{1} (\wavevec ')\conduct_{1}(\wavevec'') i {\undecomposedvec{\wavevec'}} \myphase{(\wavevec' + \wavevec'')} -
  \sum_{\wavevec} \voltage_{2}(\wavevec) \conduct_{0} i {\undecomposedvec{\wavevec}}\myphase{\wavevec}
}
\myequationn{
\meanflow_{2}(\wavevec) = \conduct_{2} \hat{z} - \voltage_{1}(\wavevec')\conduct_{1} (\wavevec'')i  {\undecomposedvec{\wavevec'}} \mysecorddelt -
 i \conduct_{0} {\undecomposedvec{\wavevec}} \voltage_{2}(\wavevec)
}

Plugging in $\voltage_{1}$, $\voltage_{2}$ gives us:

\def \jtwostuff{
\wavevec'_{z}
\frac{-\undecomposedvec{\wavevec'} \abs{\wavevec}^{2} + \wavevec \paren{\wavevec' \cdot \wavevec}  }
{\abs{\wavevec}^{2} \abs{\wavevec'}^{2}}
}

\def \jtwozstuff{
\wavevec'_{z}
\frac{-\zcomp{\wavevec'} \abs{\wavevec}^{2} + \zcomp{\wavevec} \paren{\wavevec' \cdot \wavevec}  }
{\abs{\wavevec}^{2} \abs{\wavevec'}^{2}}
}
\def \jtwoxstuff{
\wavevec'_{z}
\frac{-\xcomp{\wavevec'} \abs{\wavevec}^{2} + \xcomp{\wavevec} \paren{\wavevec' \cdot \wavevec}  }
{\abs{\wavevec}^{2} \abs{\wavevec'}^{2}}
}

\myequationn{
\meanflow_{2}(\wavevec) = \conduct_{2} \hat{z} - \voltageone{\wavevec'} \conduct_{1} (\wavevec'')i  {\undecomposedvec{\wavevec'}} \mysecorddelt 
\\
- i \conduct_{0} {\undecomposedvec{\wavevec}}\paren{\finalvolttwo}
}
We simplify this as:
\myequation{\meanflow_{2}(\wavevec) 
 = \conduct_{2}(\wavevec) \bracket{\hat{z}   \horizontality{\wavevec}  - \hat{x} \frac{\wavevec_{x}\wavevec_{z}}{\abs{\wavevec}^{2}} }  + 
\frac{\condconddelt}{\conduct_{0}} \bracket{
\jtwostuff
}.
\label{eq:secorderflow}
}

Here, we can confirm that \myequationn{
\nabla \cdot \bracket{\myphase{\wavevec} \meanflow_{2}(\wavevec) } = i  \myphase{\wavevec}  \bracket{\wavevec \cdot \meanflow_{2}(\wavevec) } = 0.
}
It will be convenient to have this answer divided into x and z components:
\myequation{\meanflow_{2}(\wavevec) 
 = \conduct_{2}(\wavevec) \bracket{\hat{z}   \horizontality{\wavevec}  - \hat{x} \frac{\wavevec_{x}\wavevec_{z}}{\abs{\wavevec}^{2}} }  + \\
\frac{\condconddelt}{\conduct_{0}} \bracket{
\hat{z} \  \jtwozstuff
+ \hat{x} \  \jtwoxstuff
} .
\label{eq:secorderflow}
}

\subsection{Change in \timederiv}

Likewise, we may represent the change in the time derivative functional:
\myequationn{
\timederiv(\dens, \meanflow, \vec{\alpha}) = \timederiv_{\dens}  \Delta \dens + \timederiv_{\meanflow} \meanflow_{1} + \timederiv_{\meanflow} \meanflow_{2} + \frac{1}{2} \paren{\timederiv_{\meanflow \meanflow} \ \meanflow_{1}^{2} + \timederiv_{\dens \dens} \ \Delta \dens^{2} + 2 \timederiv_{\meanflow \dens} \ \Delta \dens \meanflow_{1}
} + \thirdordstuff
}

\def \firstordjzderiv{
\timederiv_{\meanflow_{z}}
}

\def \secordjxderiv{
\timederiv_{\meanflow_{x} \meanflow_{x}} \xspace
}

\def \secorddensderiv{
\timederiv_{\dens\dens} \xspace
}

\def \secordjzderiv{
\timederiv_{\meanflow_{z} \meanflow_{z}} \xspace
}

\def \secordjxderiv{
\timederiv_{\meanflow_{x} \meanflow_{x}} \xspace 
}

\def \secordhybderiv{
	\timederiv_{\meanflow_{z} \dens} \xspace 
}
We note that, due to symmetry, $\timederiv$ is even with respect to $\meanflow_{x}$ are relevant. Therefore we may simplify the time derivative as:
\myequationn{
\timederiv = \timederiv_{\dens}  \Delta \dens + \firstordjzderiv \meanflow_{1z} + \firstordjzderiv \meanflow_{2z} + \frac{1}{2} \bracket{
\secordjzderiv \ \meanflow_{1z}^{2} + 
\secordjxderiv\ \meanflow_{1x}^{2} + 
 \  \timederiv_{\dens \dens} \ \Delta \dens^{2} +
 2 \secordhybderiv \ \meanflow_{1z} \Delta \dens
}
+\thirdordstuff, 
}
where we have used the notation $\meanflow_{1z} = \meanflow_{1} \cdot \hat{z}$, $\meanflow_{1x} = \meanflow_{1} \cdot \hat{x}$, $\meanflow_{2z} = \meanflow_{2} \cdot \hat{z}$.\\

We recover the first order component $\timederiv_{1}$ as before:
\myequationn{
\timederiv_{1} = \timederiv_{\dens} \Delta \dens + 
\firstordjzderiv \meanflow_{1z} \Rightarrow
\timederiv_{1}(\wavevec) =  \bracket{\timederiv_{\dens} + \firstordjzderiv \conduct_{\dens} \horizontality{\wavevec}} \dens(\wavevec).
}
We now calculate the second order component $\timederiv_{2}$:
\myequationn{
\timederiv_{2} = \firstordjzderiv \meanflow_{2z} + \frac{1}{2} \bracket{
\secordjzderiv \ \meanflow_{1z}^{2} + 
\secordjxderiv\ \meanflow_{1x}^{2} + 
 \  \timederiv_{\dens \dens} \ \Delta \dens^{2} +
 2 \secordhybderiv \ \meanflow_{1z} \Delta \dens
}
}
Which we may decompose as:
\myequationn{
\timederiv_{2}(\wavevec) = 
\firstordjzderiv \meanflow_{2z} + 
\frac{1}{2} \bracket{
\secordjzderiv \ \meanflow_{1z}(\wavevec') \meanflow_{1z}(\wavevec'')+ 
\secordjxderiv\ \meanflow_{1x}(\wavevec') \meanflow_{1x}(\wavevec'')+ 
\timederiv_{\dens \dens}  \dens(\wavevec') \dens(\wavevec') +
 2 \secordhybderiv \ \meanflow_{1z}(\wavevec')  \dens(\wavevec'')
}
\mysecorddelt
}
\def \secordjzstuff{ \horizprime{\wavevec'} \horizprime{\wavevec''}
\xspace
}
\def \secordjxstuff{ \diagonality{\wavevec'}\diagonality{\wavevec''}
\xspace 
}
\def \secordhybstuff{
\horizprime{\wavevec'}
}
\def \conddens{
\conduct_{1}(\wavevec')  \dens(\wavevec'') 
}
Substituting $\meanflow_{1z}$, $\meanflow_{2z}$, $\meanflow_{1x}$, we get:
\myequationn{
\timederiv_{2}(\wavevec) =
\firstordjzderiv 
  \bracket{ \frac{1}{2}
   \conduct_{\dens\dens} \horizontality{\wavevec}  \densdensdelt    + 
\frac{\condconddelt}{\conduct_{0}} 
 \jtwozstuff
}
+\\
\frac{1}{2} \bracket{
\secordjzderiv \  \condcond \secordjzstuff + 
\secordjxderiv\   \condcond \secordjxstuff+
\timederiv_{\dens \dens}  \densdens+
 2 \secordhybderiv \   \conddens \secordhybstuff
}
\mysecorddelt
}

We shuffle some terms around to get
\myequationn{
\timederiv_{2}(\wavevec) =
\paren{\densdensdelt}  \frac{1}{2} \bracket{\secorddensderiv +  \firstordjzderiv \conduct_{\dens\dens} \horizontality{\wavevec} }\\
+\condconddelt \bracket{ \frac{\firstordjzderiv}{\conduct_{0}} \jtwozstuff  +\frac{\secordjzderiv }{2} \paren{\secordjzstuff} + \frac{\secordjxderiv}{2} \paren{\secordjxstuff} }\\
+\conddens \mysecorddelt \secordhybderiv \secordhybstuff 
}
And substitute $\conduct_{1}(\wavevec)  = \dens(\wavevec) \conduct_{\dens}$:
\myequationn{
\timederiv_{2}(\wavevec) =
\paren{\densdensdelt}  \frac{1}{2} \bracket{\secorddensderiv +  \firstordjzderiv \conduct_{\dens\dens} \horizontality{\wavevec} }\\
+\paren{\densdensdelt }\paren{\conduct_{\dens}}^{2}  \bracket{\frac{\firstordjzderiv}{\conduct_{0}} \jtwozstuff  +\frac{\secordjzderiv }{2} \paren{\secordjzstuff} + \frac{\secordjxderiv}{2} \paren{\secordjxstuff} }\\
+\paren{\densdensdelt  } \secordhybderiv \conduct_{\dens} \secordhybstuff 
}

We obtain our final answer:
\myequationn{
\timederiv_{2}(\wavevec) = \finalinteraction(\wavevec, \wavevec', \wavevec'') \densdensdelt,
}
Where our interaction function, $\finalinteraction$, is:
\myequationn{
\finalinteraction(\wavevec, \wavevec', \wavevec'') = 
\frac{\secorddensderiv}{2}  + \frac{\firstordjzderiv \conduct_{\dens\dens} \horizontality{\wavevec}}{2}
+\frac{\firstordjzderiv \paren{\conduct_{\dens}}^{2}  }{\conduct_{0}} \jtwozstuff  \\
+\frac{\secordjzderiv \paren{\conduct_{\dens}}^{2}  }{2} \paren{\secordjzstuff} + \frac{\secordjxderiv \paren{\conduct_{\dens}}^{2}  }{2} \paren{\secordjxstuff}
+ \secordhybderiv \conduct_{\dens} \secordhybstuff .
}
We note that $\finalinteraction(\wavevec, \wavevec', \wavevec'') \neq \finalinteraction(\wavevec', \wavevec, \wavevec'')$, and thus can not be written as the derivative of some cubic free energy functional. Therefore, like the discrete model, the continuum model at second order has no detailed balance. This gives a stochastic field equation with no detailed balance and complicated interaction coefficients, and we choose not to proceed further. Note that $\eta$ will also change with $\meanflow, \dens$ with analogous first-order and second-order terms, but here, we have ignored this extra complexity.

\section{Methods}

The code used was written in a combination of C++, MATLAB, and Objective-C, and can be downloaded at
\href{http://web.mit.edu/socko/Public/PublishedCode/ActivePorousMediaCode.zip}{http://web.mit.edu/socko/Public/PublishedCode/ActivePorousMediaCode.zip}. Currents were solved using the Eigen linear equation solver \cite{eigenweb}.

  \end{document}